\documentclass[pdflatex,sn-mathphys-ay]{sn-jnl}

\usepackage{amsmath,amssymb,amsfonts}
\usepackage{amsmath,amsfonts}
\usepackage{graphicx}
\usepackage{subfig}
\usepackage{booktabs}

\newcommand*\1{\text{\usefont{U}{bbm}{m}{n}1}}

\begin{document}

\title{A model-based approach for clustering binned data}

\author*[1]{\fnm{Asael Fabian} \sur{Mart\'{i}nez}}\email{fabian@xanum.uam.mx}
\author[2]{\fnm{Carlos} \sur{D\'\i az-Avalos}}\email{carlos@sigma.iimas.unam.mx}

\affil*[1]{\orgdiv{Department of Mathematics}, \orgname{Universidad Aut\'{o}noma Metropolitana}, \orgaddress{\street{Av. San Rafael Atlixco 186}, \city{Mexico City}, \postcode{09340}, \country{Mexico}}}

\affil[2]{\orgdiv{IIMAS}, \orgname{Universidad Nacional Aut\'{o}noma de M\'{e}xico}, \orgaddress{Circuito Escolar S/N}, \city{Ciudad Universitaria}, \postcode{04510}, \country{Mexico}}

\abstract{Binned data often appears in different fields of research, and it is generated after summarizing the original data in a sequence of pairs of bins (or their midpoints) and frequencies. There may exist different reasons to only provide this summary, but more importantly, it is necessary being able to perform statistical analyses based only on it. We present a Bayesian nonparametric model for clustering applicable for binned data. Clusters are modeled via random partitions, and within them a model-based approach is assumed. Inferences are performed by a Markov chain Monte Carlo method and the complete proposal is tested using simulated and real data.
Having particular interest in studying marine populations, we analyze samples of \textit{Lobatus (Strobus) gigas}' lengths and found the presence of up to three cohorts along the year.}


\maketitle

\section{Introduction}

Data analysis often considers observations as continuous, but there are many situations where they are recorded with a very low precision during the data gathering process. This might happen since the data were measured with non high-precision instruments, due to their cost, or the lack of availability of an adequate measuring device. It could also be that the original data are only accesible in a summarized form, for simplicity or confidentiality reasons. 
In these and similar scenarios, assuming that the original observed data fall on an interval $(a,b)\subset\mathbb{R}$, they are only available after being processed as follows. First, the interval is divided into a series of non-overlapping subintervals, or bins, $T_{1}, \ldots, T_{m}$ such that their union $T_{1} \cup\cdots\cup T_{m}$ covers $(a,b)$, and then, it is counted how many datapoints fall in each bin. Each bin $T_l$ is represented by its center $\tau_l$, whose value can be its midpoint or some statistics such as the mean or median value. This data binning process produces a sequence of pairs $(\tau_1,f_1), \dots, (\tau_m,f_m)$ representing the center and frequency, or number of observations, assigned to the corresponding bin $T_l$, $l=1,\dots,m$. We call binned data to this kind of datasets.

Binned data play important roles in fields such as biology, ecology, social sciences, economy, engineering, computing, and others. For statisticians, perhaps one of the most known example of binned data is the Fisher's Iris dataset \citep{Fisher36}, consisting of $50$~samples from each of three species of Iris; each sample records five values: sepal length, sepal width, petal length, petal width and species. The variables for lengths and widths are clearly continuous but were recorded using only one decimal place. Another classic example appears in \cite{Titterington+85} as an example of finite mixture distributions, originally published by \cite{Cassie54}, where the lengths of $256$~snappers are reported as a frequency table.
In this way, many biological or ecological studies collect data using a binned structure; see, for example, \cite{Burnham+80} for line transect sampling, a method used for estimating wildlife populations, and the references in \cite{Petitgas+11} and \cite{Edwards+20} for cases of data binning in fishery.
In socio-economic studies, data related to income or poverty are mainly available in binned form for developing countries, or due to the high cost of sharing the raw data (see \citealt{Hajargasht+12} and \citealt{Eckernkemper+21} among others); similarly for schooling and child welfare (see, \citealt{Guo05} and \citealt{Friedman+19}). It also happen that organizations wish to protect data confidentiality, security, and integrity, as explained by \cite{Zahra+22}, so only binned data are made public.

On a different direction, there are cases where raw data are binned in order to avoid computational burdens, or to simplify further processing, like in \cite{Hamdan06} and \cite{Same+06}. Also, the process of binning has been used to filter some issues associated to the data-generating phenomena \citep{Hecht+23}. Finally, data might be originally collected in a binned form, like in survival analysis, where it is common to have interval-censored data, or in pattern recognition and machine learning problems, where color and feature histograms are the input data (see \citealt{Cadez+02} and the references therein for more details).

Although descriptive analysis or some elemental estimations might be performed using the binned data as is, these can be inaccurate by some kind of bias inherent to the data, and aiming to perform more elaborated or specific statistical analyses become more cumbersome. Among the literature, a wide range of methodologies have been proposed for the statistical treatment of binned data. For example, in parametric estimation, there are methods based on maximum likelihood \citep{Tarsitano05,Edwards+20,Eckernkemper+21}, in the method of moments \citep{Hajargasht+12}, or following Bayesian approaches \citep{Eckernkemper+21}. When the data exhibit a more complex distribution, kernel density estimation proposals are widely used; they can be found, for example, in \cite{Scott81}, \cite{Scott+85}, \cite{Blower+02}, \cite{Cadez+02}, \cite{Jang+10}, \cite{WW13}, \cite{Reyes+16}, \cite{Xiao16}, and \cite{Hecht+23}. A different approach is given by \cite{Koo+00}. 
Another popular class of methods along this line is based on a modification of the EM algorithm \citep{D+77} to handle missing data. Two extensions are given by \cite{McLachan+88} and \cite{CD88}, and have been used in mixture modeling by \cite{C95}, \cite{Hamdan06}, \cite{Same+06} and \cite{Zahra+22}. 
With regard to fully Bayesian methods designed for binned data, \cite{LE09} propose a density estimation model, and \cite{AM10} and \cite{Gau+14} estimate the parameters for finite mixture models.

Motivated by the study of marine species in Quintana Roo coasts, we develop a Bayesian nonparametric methodology for estimating the underlying clustering structure based on collected data, available only in a binned form.
Providing a clustering structure may allow to identify different cohorts in the population, and estimate some characteristics of interest. This information can subsequently be incorporated in growth models, a common tools for fisheries management (see, for example, \citealt{Lorenzen16}).

Our proposed methodology, on the one hand, models heterogeneity in the data following a clustering perspective by means of random partitions \citep{Ewens72,Kingman78g,Kingman78p}, in particular, a modification of the distribution over partitions for the Dirichlet process \citep{Ferguson73} is used. On the other hand, data within groups are modeled under a probabilistic framework, also known as a model-based approach \citep{FR02}. However, since the data is provided in a binned form, it is necessary to extend this framework; which we tackle by treating original data as missing values.
Once the underlying clustering structure is estimated, we can easily estimate group's parameters, as well as some other characteristics, like the distribution of the data. We test the performance of our methodology using simulated and real data. Finally, we analyze samples of the length of \textit{Lobatus (Strombus) gigas}, collected at different times in Punta Gavil\'an, Quintana Roo.

\section{Random partition models}

For any dataset $y=\{y_1,\dots,y_n\}$ of $n$~elements, cluster analysis, also known as clustering, aims to provide an estimate of their grouping structure, say $\pi$, which partitions $y$ into nonempty subsets $\pi_j$, $j=1,\dots,k$. Elements belonging to the same subset, or group in the context of clustering, have more similar characteristics among them than those in any other group. Given the importance of the task of clustering, many methodologies have been developed. Some of them cluster the elements according to their distance, in some sense, to the other elements in the same group or to some representative of the group; hierarchical clustering and the $k$~means algorithm are examples of these methods. Other approaches make use of a probabilistic component to associate the elements to a single group. There, it is assumed that each element is a realization of some random variable, so all elements belonging to the same group $\pi_j$ were generated from the same probability distribution $g_j$; this approach is known as model-based \citep{FR02}. A common example of model-based methods are mixture models, i.e. convex linear combinations of probability distributions.

A different probabilistic approach for cluster analysis is based on random partitions. Early works on this topic can be traced back to \cite{Ewens72} and \cite{Kingman78g,Kingman78p} for the study of population genetics, and, nowadays, they are widely used in Bayesian nonparametric models; see, e.g.\ \cite{Hjort+10} and \cite{DeBlasi+15} for a more extensive overview. A random partition is a random object taking values over the space generated by all the possible groupings of the $n$~elements, i.e.\ every possible value of $\pi$. When they are combined with the model-based approach, we can determine the grouping structure underlying to a given dataset. By providing a probabilistic framework for all the possible groupings, we can have \textit{a~priori} and \textit{a~posteriori} probabilities for each possible value of $\pi$, as well as some other estimates of interest. An example of this class of models is known as product partition models \citep{Hartigan90}. Throughout this work, we refer this approach for clustering as random partition models.

A natural way to describe random partition models is through Bayesian hierarchical models. Let us denote by $\mathcal P$ the space of all the possible values of $\pi$, in terms of the observations' indices, and assume that the probability distribution for each group, $g_j$, also called kernel, is the same and only its parameter, $\theta_j$, varies, so $g_j(\cdot):=g(\cdot\mid\theta_j)$. Then, the hierarchical model can be written as
\begin{equation}
\label{eq:partition.model}
\begin{aligned}
  y_i\mid\theta,\pi & \sim g(y_i\mid\theta_j)\1(i\in\pi_j),\text{\ [ind]}\quad i=1,\dots,n\\
  \theta_j\mid\pi & \sim \nu_0(\theta_j),\text{\ [iid]}\quad j=1,\dots,k\\
  \pi &\sim \rho_0(\pi),
\end{aligned}
\end{equation}
where $\nu_0$ and $\rho_0$ are prior distributions for the kernel parameters and the random partition, respectively.

The prior $\nu_0$ in Model~\eqref{eq:partition.model} can be chosen such that, together with $g$, both form a conjugate model. On the other hand, the prior for the random partition $\pi$, $\rho_0$ seems less straightforward. For example, \cite{Hartigan90} defines partition distributions as proportional to the product of non-negative \textit{cohesion} functions, $\kappa$, applied to each group in $\pi$, i.e.
\begin{equation*}
\Pr(\pi=\pi_1/\cdots/\pi_k)\propto\prod_{j=1}^{k}\kappa(\pi_j),
\end{equation*}
for all partition $\pi_1/\cdots/\pi_k:=\{\pi_1,\dots,\pi_k\}\in\mathcal{P}$.

A different class of partition distributions arises when working with discrete almost surely (a.s.) random probability measures (RPMs), daily used tools in Bayesian nonparametrics; see, for instance, \cite{Regazzini+03}, \cite{IJ01}. Any sample of size~$n$ from an a.s.\ discrete RPM $\tilde p$, taking values in $\mathbb{X}$, is bounded to have ties with positive probability, and, as consequence, to form $K=k$ groups with distinct representative values, each one having frequency $N_1,\dots,N_k$, with $1\leq k\leq n$. Therefore, a partition distribution is obtained through
\begin{equation*}
    p(n_1,\dots,n_k)=\Pr(\{K=k\}\cap\{N_1=n_1,\dots,N_k=n_k\})=\int_{\mathbb{X}^k} \mathbb{E}\left(\prod_{j=1}^k\tilde p^{n_j}(\mathrm{d}x_j)\right),
\end{equation*}
by letting $\Pr(\pi=\pi_1/\cdots/\pi_k)=p(\#\pi_1,\dots,\#\pi_k)$, with $\# A$ the cardinality of $A$. In the computation of the expression above, the expectation is with respect to the distribution of $\tilde p$. The resulting class of distributions is known as exchangeable partition probability functions (EPPFs), where exchangeable stands for symmetric since it can be noticed that the probability is the same under permutations of the groups. Further, \textit{a priori}, it does not matter what the sampled values of $\tilde p$ are in order to compute the probabilities, and more importantly, it can be seen that EPPF-based models also allow us to infer about the number of groups, denoted by the random variable $K$. The Dirichlet process \citep{Ferguson73}, the daily workhorse in Bayesian nonparametrics, is the canonical example for the distribution of $\tilde p$, and its EPPF is given by
\begin{equation}\label{eq:eppf}
  p(n_1,\dots,n_k)= \frac{\alpha^k}{(\alpha)_n}\prod_{j=1}^k\Gamma(n_j),
\end{equation}
for some $\alpha>0$, where $n=n_1+\cdots+n_k$, and $(x)_n=\Gamma(x+n)/\Gamma(x)$ is the Pochhammer symbol.
More general constructions can be found in, for example, \cite{Lijoi+07biom,Lijoi+07jrss}, \cite{Favaro+16} and \cite{GilMena23}.

\subsection{Modeling binned data}

Model~\eqref{eq:partition.model} is designed to be used when the observations $y$ are all available, that is when each $y_i$ has been recorded with enough precision to be considered as a realization of some continuous random variable. However, if data are only available in a binned form, an additional treatment is required.

As defined in the Introduction, binned data are obtained after partitioning an interval $(a,b)$, where the original observations fall, into $m$~non-overlapping subintervals $T1,\dots,T_m$, and recording the number of observations, $f_l$, falling within each bin. Therefore, variables $y_1,\dots,y_n$ are non observable, and in order to still following a model-based approach, the observation-level hierarchy in Model~\eqref{eq:partition.model} models a series of latent variables, and it needs to be modified as
\begin{equation}\label{eq:latent}
  y_i\mid\theta,\pi,T,e \sim g(y_i\mid\theta_j)\1(i\in\pi_j)\1(y_i\in T_{e_i}),\text{\ [ind]}\quad i=1,\dots,n,
\end{equation}
where $e_i$ is an indicator variable associating each observation $y_i$ to the $e_i$th~bin; so $\sum_{i=1}^n\1(e_i=l)=f_l$, for $l=1,\dots,m$, must be satisfied. This is a common approach in Bayesian methods dealing with missing values.

\subsection{Sampling method}

Making inferences from random partition models, like~\eqref{eq:partition.model}, can be done using some sampling schemes designed for mixture models; in our case, for the Dirichlet process mixture model (see, \citealt{EW95}, \citealt{Neal00} or \citealt{IJ01}, just to mention a few examples). In some of them, a set of indicator variables associating observations and mixture components is introduced to make inferences, and there is a close relationship between this set and a random partition; a detailed discussion about it can be found in \cite{GM23}. On a different direction, \cite{FMW10} propose a sampling method specifically designed for clustering, also based on the Dirichlet process mixture model.

Our model for binned data can use any of the mentioned sampling methods after adding the extra step given in Equation~\eqref{eq:latent} for sampling the latent observations $y$. We adopt the method of \cite{FMW10} since it simplifies the exploration of the whole space $\mathcal P$, where random partitions take values.
Their approach is designed to cluster univariate observations by first ordering them. If observations $y_{j}$ and $y_{l}$ are clustered together, with $y_{j}<y_{l}$, it is reasonable to assume that all observations in between should belong to the same cluster. As a consequence, the possible groupings are reduced from the $n$th Bell number of choices, the cardinality of $\mathcal P$, to $2^{n-1}$, with $n$ the sample size. See \cite{FMW10} and \cite{MM14} for further details. We denote by $\mathcal S$ this reduced space of possible groupings. After a partition is simulated, additional parameters, like those associated to each group, can be also sampled.

Taking the EPPF for the Dirichlet process, Equation~\eqref{eq:eppf}, its restriction to $\mathcal S$ leads to the following expression
\begin{equation*}
\rho_0(\pi_1/\cdots/\pi_k)= \binom{n}{\#\pi_1,\dots,\#\pi_k}\frac{1}{k!}p(\#\pi_1,\dots,\#\pi_k).
\end{equation*}
Furthermore, regarding the bin-membership variables $e_1,\dots,e_n$, they take the following particular structure under this restriction
\begin{gather*}
e_1=e_2=\dots=e_{f'_1}=1\\
e_{f'_1+1}=e_{f'_1+2}=\dots=e_{f'_2}=2\\
\vdots\\
e_{f'_{m-1}+1}=e_{f'_{m-1}+2}=\dots=e_{f'_m}=m.
\end{gather*}
with $f'_l=f_1+\cdots+f_l$.

Therefore, we require to compute the posterior distribution of variables $y$, $\theta$, and $\pi$, which are given by
\begin{equation*}
p(y,\theta,\pi\mid T,e)\propto p(y\mid\theta,\pi,T,e) p(\theta\mid\pi,T,e) p(\pi\mid T,e).
\end{equation*}
Their corresponding full conditional distributions are given as follows. For the latent variables $y_i$, $i=1,\dots,n$,
\begin{equation*}
p(y_i\mid y_{-i},\theta,\pi,T,e) \propto g(y_i\mid\theta_j)\1(i\in\pi_j)\1(y_i\in T_{e_i}),
\end{equation*}
which is a $T_{e_i}$-truncated version of $g$ with parameter $\theta_j$. The notation $y_{-i}$ indicates the vector $(y_1,\dots,y_n)$ without its $i$th entry. Further, the posterior for the kernel parameters $\theta_j$, $j=1,\dots,k$, is
\begin{equation*}
p(\theta_j\mid\theta_{-j},\pi,T,e)\propto \nu_0(\theta_j) \prod_{i\in\pi_j}g(y_i\mid\theta_j);
\end{equation*}
sampling from this distribution can be straightforward, for example using conjugate models.

Finally, there is the distribution for the $\mathcal S$-valued random partition, from which we can sample according to the details provided by \cite{FMW10}. Briefly, the Markov chain Monte Carlo (MCMC) scheme to draw samples for $\pi$ consists on the following steps. Assume the current value of $\pi$ has $k$~groups, and let $\pi'$ be a proposed value; then
\begin{enumerate}
  \item the partition $\pi'$ will have $k+1$~groups, where some non-singleton group $\pi_j$ was chosen and split into two groups;
  \item conversely, the partition $\pi'$ will have $k-1$~groups, where two adjacent groups of $\pi$ were chosen and merged.
\end{enumerate}
Each move is a Metropolis-Hastings step and is accepted accordingly. After a split/merge is performed, an additional step is done by updating the size of two adjacent groups, the so-called shuffle move. The details of each move for this scheme are provided in the Appendix of \cite{MM14}.

It is worth mentioning that our main interest is to infer about the clustering structure underlying the data, which corresponds to a point estimate for parameter $\pi$, denoted by $\hat\pi$. Despite of the different estimators available in the literature (see, e.g.\ \citealt{Dahl06} and \citealt{Wade+18}), we report the posterior modal partition due to its simplicity. Furthermore, due to the ordering restriction in the possible groupings, we will report partitions in terms of their groups' sizes, for instance, $\pi=(3,2,4)$ means there are three groups, in terms of the indices: $\{1,2,3\}$, $\{4,5\}$, and $\{6,7,8,9,10\}$.

Once the point estimate $\hat\pi$ is computed, the estimation of kernel parameters can be done by taking all sampled values $(\theta_1,\dots,\theta_k)^{(t)}$, at iteration~$t$, where the sampled partition $\pi^{(t)}=\hat\pi$, that is
\begin{equation*}
    \hat\theta_j=\frac{1}{N(\hat\pi)}\sum_{t=1}^{T}\theta_j^{(t)}\1(\pi^{(t)}=\hat\pi),\quad j=1,\dots,k,
\end{equation*}
where $N(\hat\pi)=\sum_{t=1}^T\1(\pi^{(t)}=\hat\pi)$, and $T$ is the number of sampled values. Notice that following this approach, and taking advantage of the restriction induced by $\mathcal{S}$, there is no label switching issues; each individual sampled $\theta_j$ models the $j$th~group. Moreover, the estimated density can be also computed conditioned on $\hat\pi$ as
\begin{equation*}
    \hat{f}(y\mid\hat\pi)=\frac{1}{N(\hat\pi)}\sum_{t=1}^{T}\Bigg(\sum_{j=1}^{k^{(t)}} \frac{n^{(t)}_j}{n}g(y\mid\theta^{(t)}_j)\Bigg)\1(\pi^{(t)}=\hat\pi),
\end{equation*}
with $n^{(t)}_j$ the size of group $\pi^{(t)}_j$ and $k^{(t)}$ the number of groups in $\pi^{(t)}$, leading a simple and more interpretable mixture density model.

\section{Model performance}
\label{sec:simulation}

We illustrate the performance of our model by using some datasets. In these cases, a normal density is set for modeling individual groups, i.e.\ $g$, with unknown mean and precision, so $\theta_j=(\mu_j,\lambda_j)$, and a conjugate normal-gamma prior $\nu_0$ is chosen. Also, the restricted EPPF for the Dirichlet process is set for the prior over partitions, $\rho_0$. Therefore, the model is
\begin{equation*}
\begin{aligned}
y_i\mid\mu,\lambda,\pi &\sim N(y_i\mid\mu_j,\lambda_j^{-1})\1(i\in\pi_j)\1(y_i\in T_{e_i}),\quad i=1,\dots,n\\
\mu_j,\lambda_j\mid\pi &\sim N(\mu_j\mid\omega,c/\lambda_j)Ga(\lambda_j\mid a,b),\quad j=1,\dots,k\\
p(\pi\mid\alpha) &= \frac{n!}{k!}\frac{\alpha^k}{(\alpha)_{n}}\prod_{j=1}^k\frac{1}{\#\pi_j},
\end{aligned}
\end{equation*}

The full conditional distribution for the latent variables $y_i$ are, then, a $T_{e_i}$-truncated normal distribution of parameters $(\mu_j,\lambda_j)$. For the kernel parameters $(\mu_j,\lambda_j)$, we have a normal-gamma distribution of parameters
\begin{gather*}
\omega'_j=\frac{cn_j\bar{y}_j+\omega}{cn_j+1},\qquad
c'_j=\frac{c}{cn_j+1},\\
a'_j=a+\frac{n_j}{2},\qquad
b'_j=b+\frac12\sum_{i\in\pi_j}(y_i-\bar{y}_j)^2+\frac{n_j(\bar{y}_j-\omega_j)^2}{2(cn_j+1)},
\end{gather*}
with $\bar{y}_j=\frac1{n_j}\sum_{i\in\pi_j}y_i$.
Additionally, a gamma prior is assigned to the total mass parameter $\alpha$ for the prior distribution of $\pi$; its posterior is a mixture of gamma distributions as explained in \cite{EW95}.

It is important to say that in many applications, the bins $T_l$, for $l=1,\dots,m$, are not provided, but only a representative value $\tau_l$ as explained in the Introduction. Without any further details, we assume the bins take the form $T_l=(t_{l-1},t_l]$, with $t_l=\tau_l+(\tau_{l+1}-\tau_{l})/2$, for $l=1,\dots,m-1$, $t_0=\tau_1-(\tau_{2}-\tau_{1})/2$, and $t_m=\tau_m+(\tau_{m}-\tau_{m-1})/2$.

\begin{figure}
  \centering
  \includegraphics[scale=0.5]{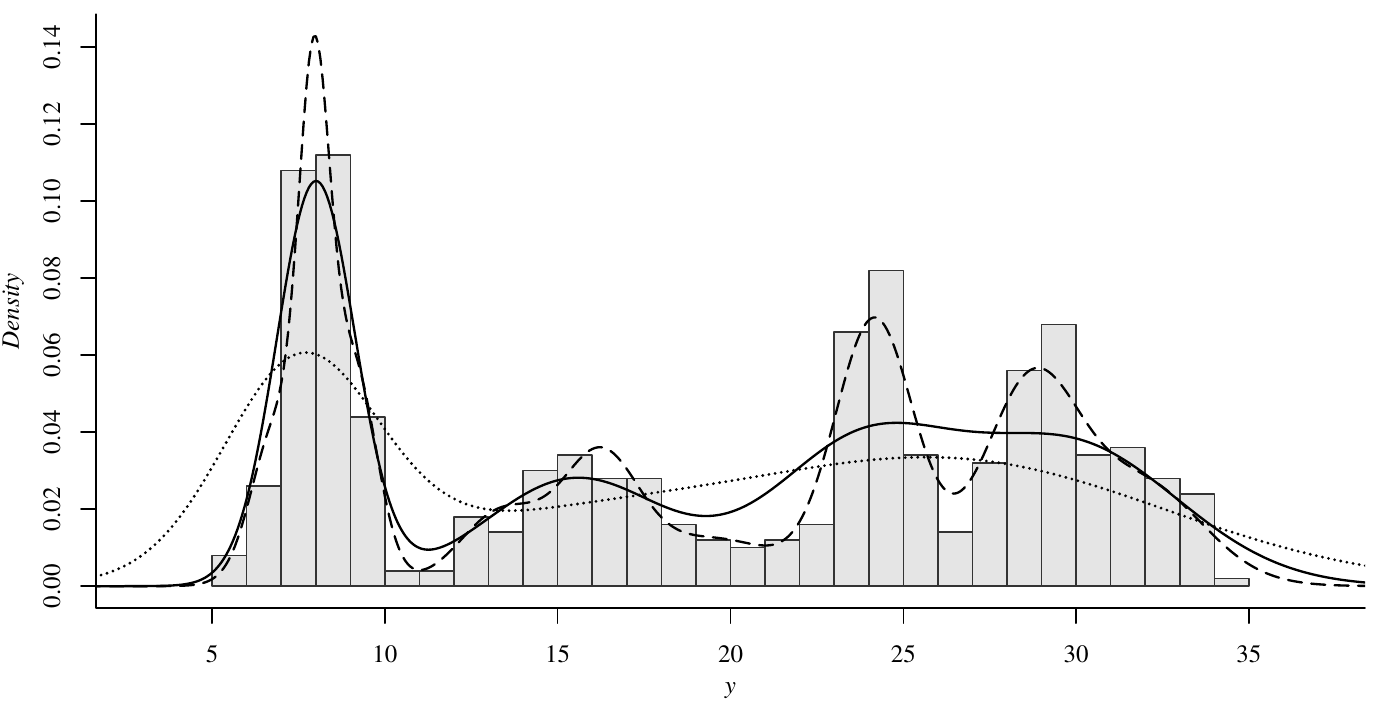}
  \caption{Density histogram for the simulated dataset from a mixture of normal distributions, and estimated densities for three values of parameter $c$: $1$ (continuous line), $0.1$ (dotted line) and $10$ (dashed line).}
  \label{fig:mixture}
\end{figure}

As a first example, a synthetic dataset is used. A sample of size $n=500$ is drawn from the following mixture of normal distributions
\begin{equation*}
  f(y)= 0.3\text{N}(y\mid8,1) + 0.2\text{N}(y\mid16,6) + 0.2\text{N}(y\mid24,1) + 0.3\text{N}(y\mid30,4),
\end{equation*}
where the second parameter of the densities is expressed in terms of the variance. Given the sample, data was binned taking the interval $(5,35)$, and $m=30$~bins where obtained, each one of length one, together with their frequencies. The histogram is depicted in Figure~\ref{fig:mixture}. The MCMC was run $30,000$~iterations long, discarding the firsts $20,000$. Base measure parameters were set to $(\omega,a,b)=(0,1.1,1)$ and different values for $c$ were used, in particular $1$, $0.1$, and $10$. Hyperparameters for the total mass parameter $\alpha$ are set to $(1,1.1)$. For each configuration, the posterior modal partition $\hat\pi$ was computed, and conditioned on it, the density and groups' parameters where also obtained. Figure~\ref{fig:mixture} also displays the estimated density for each case.

When parameter $c=1$, the estimated partition is $\hat\pi=(150,92,117,141)$, from which the mixing weights can be estimated by dividing each group size by the sample size, being them $(0.300,0.184,0.234,0.282)$. Moreover, conditioned on $\hat\pi$, the estimated means are $(8.00, 15.51, 23.79, 29.91)$, and the standard deviations, $(1.14, 2.63, 2.63, 3.11)$; the estimated density is the continuous line in Figure~\ref{fig:mixture}.

The other values for parameter $c$ allow us to see how it acts as a smoothing parameter. When $c=0.1$, the posterior modal partition contains a few groups, three, and the estimated density is coarser or softer (dotted line in Figure~\ref{fig:mixture}) than the case $c=1$. On the other hand, more groups are detected when taking $c=10$, the posterior mode contains nine groups, and the estimated density follows closer all bin heights (dashed line in Figure~\ref{fig:mixture}).

\begin{figure}
  \centering
  \includegraphics[scale=0.5]{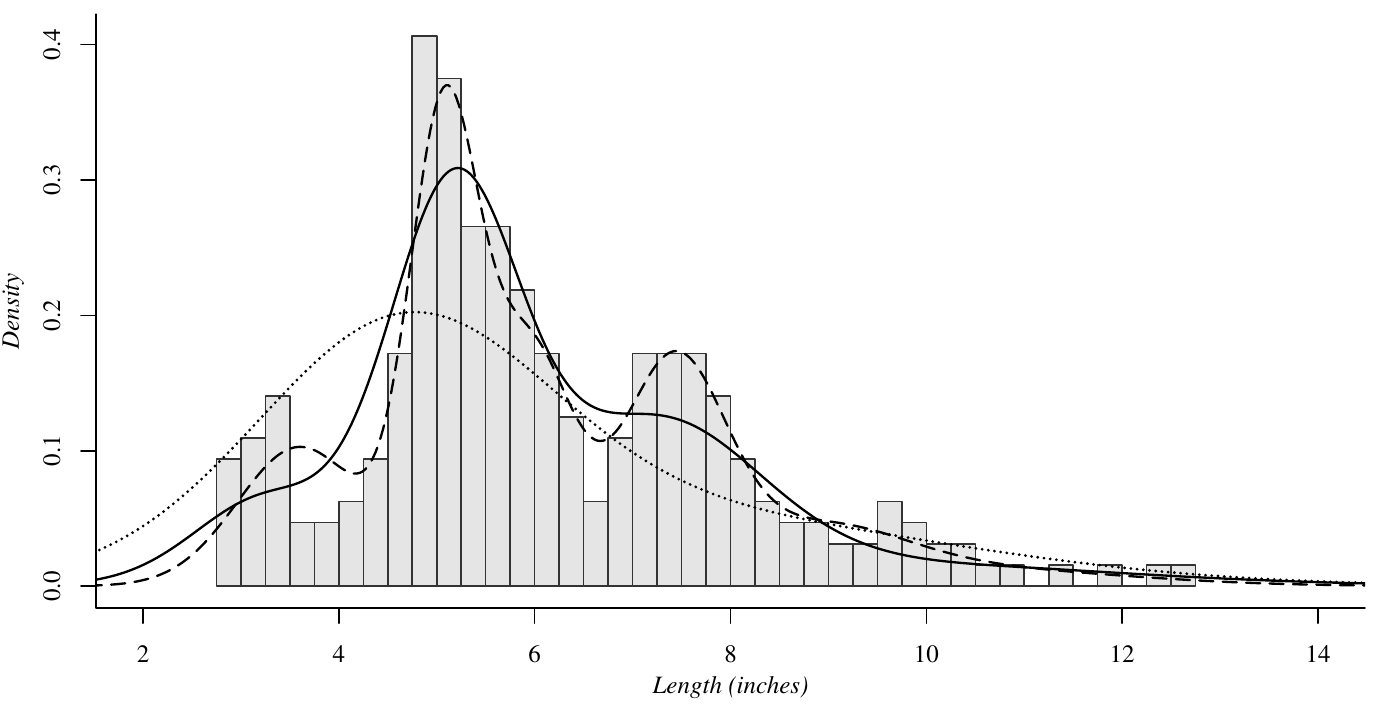}
  \caption{Density histogram for the snapper dataset of \cite{Cassie54}, and estimated densities for three values of parameter $c$: $1$ (continuous line), $0.1$ (dotted line) and $10$ (dashed line).}
  \label{fig:snapper}
\end{figure}

The second dataset used corresponds to the measurements of the length, in inches, of $256$~snappers \citep{Cassie54}, referred in the Introduction. Figure~\ref{fig:snapper} presents the histogram. The same three sampling schemes as before were used, and a similar behaviour is observed when the value of parameter $c$ varies. For the case $c=1$, four groups are estimated, whose means are $(3.24, 5.17, 7.24, 9.72)$. It is worth mentioning that these estimates are similar to those presented in \cite{Titterington+85}.

\section{Cluster analysis of \textit{Lobatus (Strombus) gigas}' length samples}
\label{sec:real}

\begin{figure}
\centering
\subfloat[][March 2010]{\includegraphics[width=0.47\linewidth]{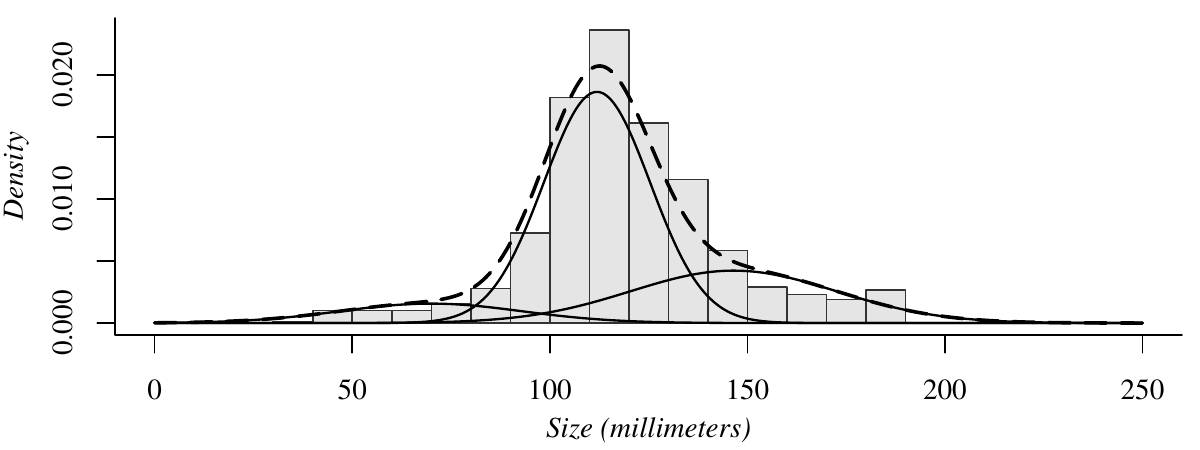}}\hfill
\subfloat[][April 2010]{\includegraphics[width=0.47\linewidth]{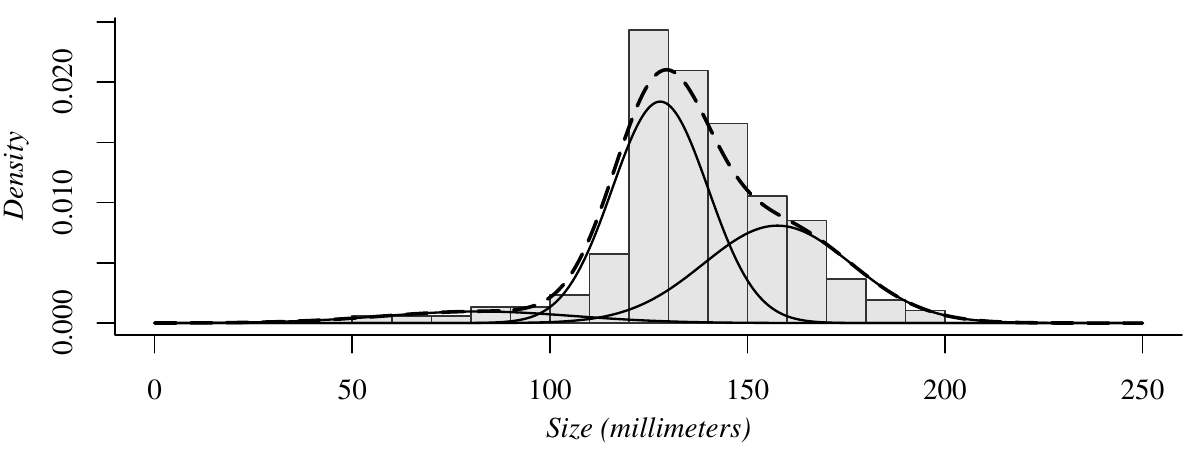}}\par
\subfloat[][July 2010]{\includegraphics[width=0.47\linewidth]{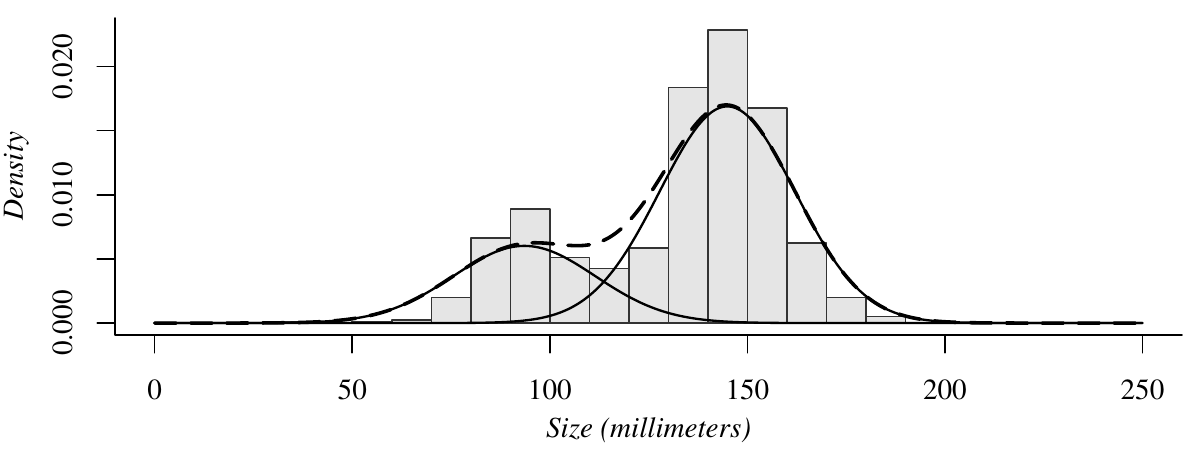}}\hfill
\subfloat[][August 2010]{\includegraphics[width=0.47\linewidth]{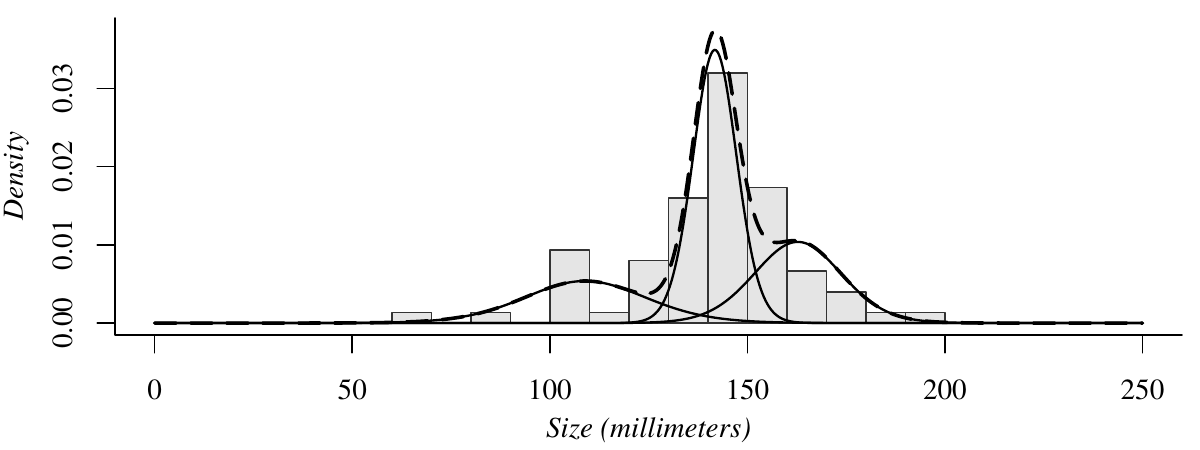}}\par
\subfloat[][September 2010]{\includegraphics[width=0.47\linewidth]{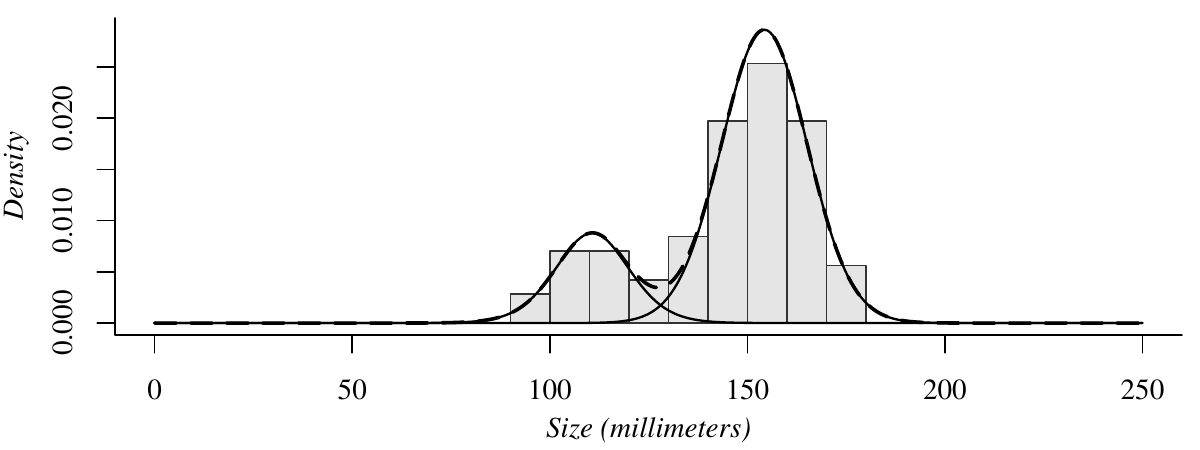}}\hfill
\subfloat[][November 2010]{\includegraphics[width=0.47\linewidth]{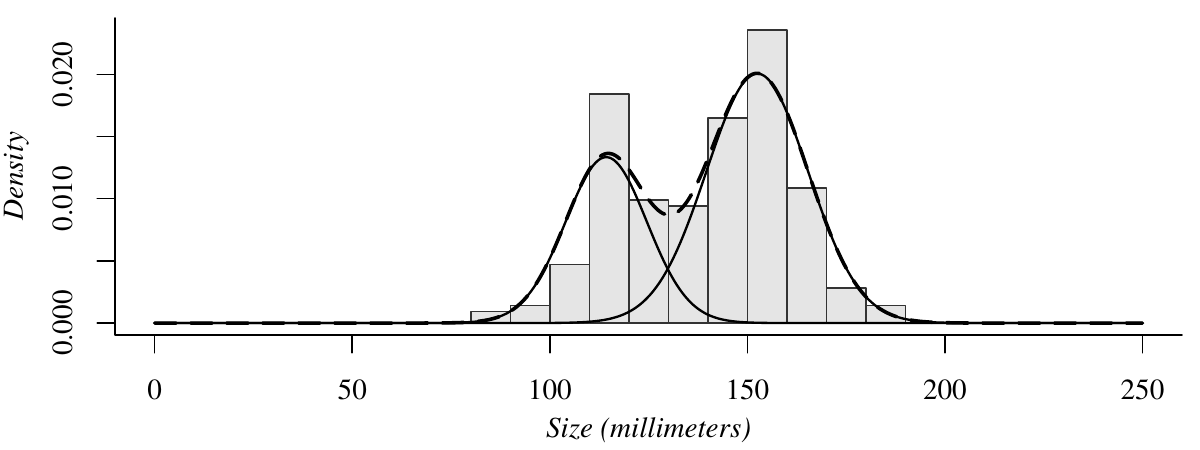}}\par
\subfloat[][December 2010]{\includegraphics[width=0.47\linewidth]{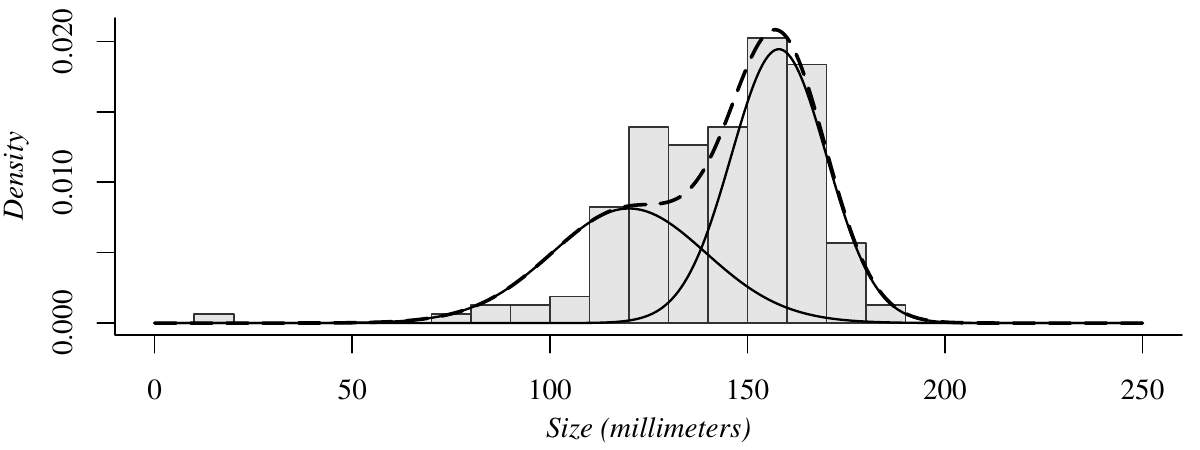}}\hfill
\subfloat[][January 2011]{\includegraphics[width=0.47\linewidth]{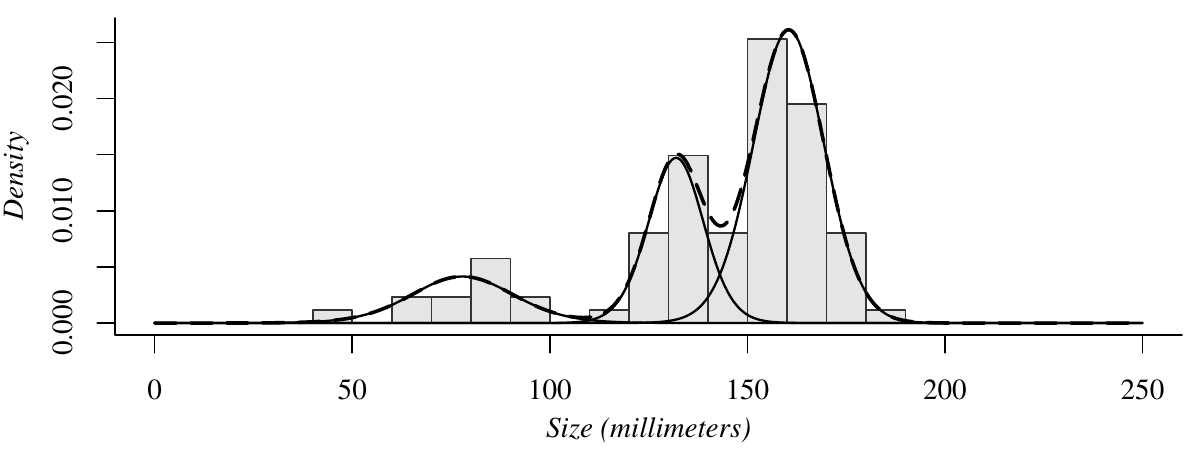}}\par
\subfloat[][February 2011]{\includegraphics[width=0.47\linewidth]{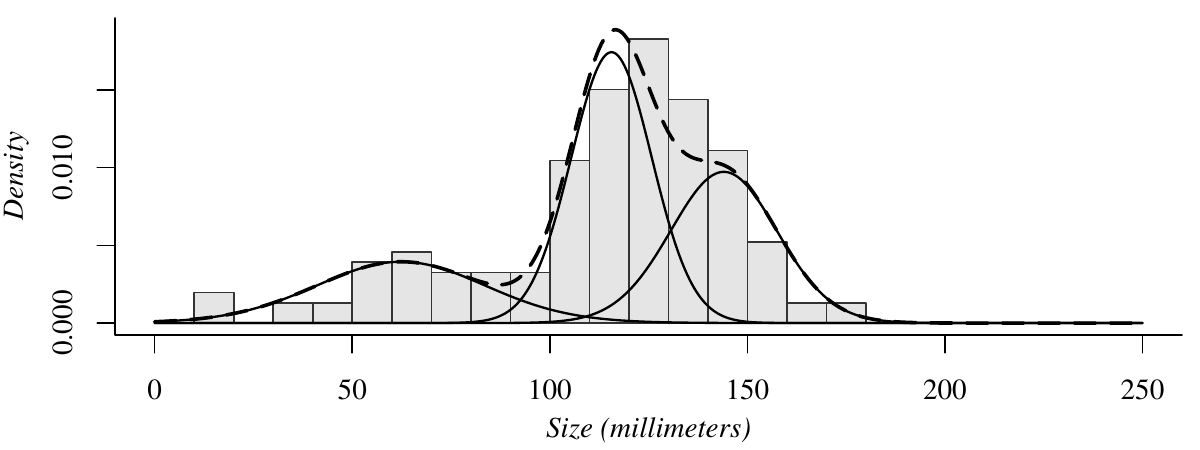}}\hfill
\subfloat[][March 2011]{\includegraphics[width=0.47\linewidth]{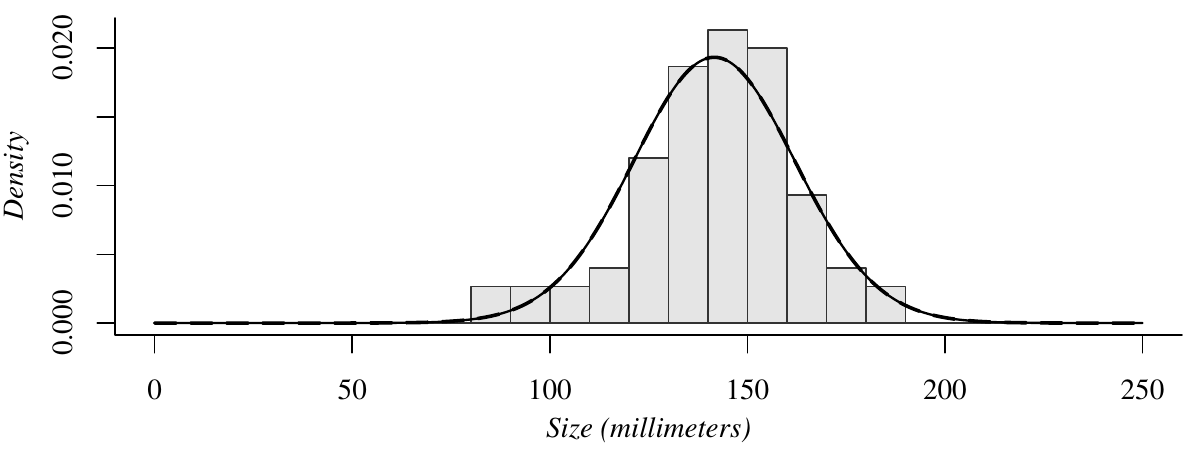}}\par
    \caption{Histograms and estimated densities for \textit{L.\ gigas}' lengths, in millimeters, recorded at different times. In all cases, bins have a width of $10$~mm. Continuous lines represent the posterior modal density estimators, and the dashed lines correspond to their mixture components.}
    \label{fig:seashell}
\end{figure}

Studies of population demographics of marine resources are important for the management and to guarantee the sustainability of exploited marine populations. Fisheries management requires high quality estimates of length- and weight-at-age to incorporate such information in harvest and optimal management models (see, for example, \citealt{Petitgas+11} and \citealt{Ogle+17}). Growth models commonly use length-at-age relationships obtained from modal progression analysis; see, e.g.\ \cite{Dippold+16}. Most bony fishes produce growth rings in their scales, which can be used to obtain length-at-age keys. Other marine resources, such as molluscs and shellfishes, do not produce growth rings and the usual way to obtain the length-at-age keys is through the analysis of modes from histograms constructed from samples from commercial catches or from field surveys. Estimating the number and the location of those modes is not always trivial due to the existence of size overlapping from different cohorts. In this section, we apply our proposed model to analyze the modal progression of \textit{Lobatus (Strombus) gigas} using data collected at 10~points in time in Punta Gavil\'an, Quintana Roo, M\'exico. The data were binned using 10-millimeter intervals (Figure~\ref{fig:seashell}).

The method was run for each sample, one by date, taking $10\,000$~iterations after discarding a first lot of~$20\,000$. Base measure parameters were fixed $(\omega,a,b)=(0,1.1,1)$; regarding parameter $c$, it was found that it has some relationship with the sample size, so a different value was chosen for each one. Figure~\ref{fig:seashell} displays, for each sample, the modal estimated density (dashed line) together with its respective mixture components (continuous lines); Table~\ref{tab:seashell} contains the estimated modal partition, and estimated posterior means and standard deviations given this partition.

\begin{table}
\centering
\begin{tabular}{l*3r}\toprule
\multicolumn1c{\itshape Date} & \multicolumn1c{\itshape Group size} & \multicolumn1c{\itshape Mean} & \multicolumn1c{\itshape Std. dev.} \\ \midrule
March 2010 & $72$ & $69.88$ & $23.29$ \\
 & $496$ & $111.94$ & $13.50$ \\
 & $218$ & $146.49$ & $26.18$ \\ \addlinespace
April 2010 & $40$ & $82.05$ & $24.24$ \\
 & $382$ & $127.95$ & $12.15$ \\
 & $260$ & $157.59$ & $18.82$ \\ \addlinespace
July 2010 & $218$ & $93.54$ & $18.07$ \\
 & $582$ & $144.93$ & $17.18$ \\ \addlinespace
August 2010 & $16$ & $109.06$ & $15.83$ \\
 & $37$ & $141.77$ & $5.62$ \\
 & $22$ & $162.78$ & $11.27$ \\ \addlinespace
September 2010 & $15$ & $110.71$ & $9.65$ \\
 & $56$ & $154.33$ & $10.99$ \\ \addlinespace
November 2010 & $74$ & $114.25$ & $10.45$ \\
 & $138$ & $152.48$ & $12.94$ \\ \addlinespace
December 2010 & $64$ & $120.02$ & $19.86$ \\
 & $94$ & $157.96$ & $12.20$ \\ \addlinespace
January 2011 & $12$ & $77.97$ & $13.20$ \\
 & $23$ & $131.90$ & $7.17$ \\
 & $52$ & $160.41$ & $9.19$ \\ \addlinespace
February 2011 & $33$ & $62.26$ & $21.83$ \\
 & $69$ & $115.60$ & $10.29$ \\
 & $51$ & $144.02$ & $13.70$ \\ \addlinespace
March 2011 & $75$ & $141.60$ & $20.66$ \\ \bottomrule
\end{tabular}
\caption{Point estimates for the \textit{L.\ gigas} data at each time: posterior modal partition, in terms of groups' sizes, mean and standard deviation for each mixture component.}
\label{tab:seashell}
\end{table}

The histograms and the estimated densities for each sample, as presented in Figure~\ref{fig:seashell}, show the presence of at least two groups (modes).
It is believed that each mode corresponds to what is called a ``year-class'' in fisheries. The continuous growth process of \textit{L.\ gigas} can be inferred from the displacement of the modal peaks to the right of the $x$-axis. The posterior means in Table~\ref{tab:seashell} show that for example, the mode at $69.88$~mm in March~2010 moved to $82.05$~mm in April~2010 and reached a value of $144.02$~mm in February~2011. A similar pattern can be tracked for other modes. 
Moreover, in March and September~2010, and in February~2011, a third mode appears at the left side of the histogram, which corresponds to the incorporation of a new cohort to the study area.

The modal progression obtained from the model we propose can also be incorporated in growth models, such as those by \cite{Gompertz25} and \cite{Bertalanffy34} (cf.\ \citealt{Hernandez+04} and \citealt{Lorenzen16}). The resulting growth rates and parameters can be incorporated to stock and management assessment process to help resource managers to maintain a sustainable fishery for \textit{L.\ gigas}. The method proposed here has the advantage of detecting the number of modal peaks and their location in an automatic way, unlike methods like \cite{Bhattacharya67}, where the user has to detect the modes visually and guess their location. Other methods like kernel density estimation, the user relies on additional procedures for detecting the modal peaks.

\section{Final remarks}

The statistical analysis of binned data is of interest in different applied fields. We have presented a Bayesian nonparametric approach aiming, for univariate binned data exhibiting a complex structure, to perform cluster analysis together with parameter estimation for detected groups.
In particular, we applied it to analyze samples of the length of \textit{L.\ gigas}, detecting cohorts and their numerical characteristics.

Under a Bayesian perspective, the absence of the original observations was handled by a simple data augmentation mechanism for missing data, that is by incorporating an extra level in the hierarchical model. This approach places our method in the model-based paradigm, which allows to provide an estimated model for the individual groups. Furthermore, it is also possible to infer about the complete model, meaning using the complete mixture model, for example, through its estimated density.
Even more, the usage of random partitions in the modeling let us to infer about the number of groups, a feature hardly available in other methodologies. Computationally, restricting their support to only consider clusters with no gaps, eases the exploration of the whole space of possible groupings during the posterior sampling.

The results obtained for the real dataset, \textit{L.\ gigas}' length, allows to understand the composition of the species at different times of the year. As explained, besides of automatically estimating the number of cohorts, our method do not require any extra procedure for detecting groups' numeric characteristics, like the mean value. Also, all this information can be further explored by incorporating it in growth models, or similar tools; their results might be useful for resource management.

Finally, our model can be extended in several ways. It is straightforward handling non-homogeneous bins, i.e.\ bins having different lengths, or using another univariate data model for groups. Moreover, the extension to analyze data in higher dimensions mainly requires an appropriate sampling scheme of the truncated data distribution.

\bibliography{myrefs.bib} 

\end{document}